\documentclass[lettersize,journal,9pt]{IEEEtran}
\usepackage{amsmath,amsfonts}
\usepackage{algorithmic}
\usepackage{algorithm}
\usepackage{array}
\usepackage[caption=false,font=normalsize,labelfont=sf,textfont=sf]{subfig}
\usepackage{textcomp}
\usepackage{stfloats}
\usepackage{url}
\usepackage{verbatim}
\usepackage{graphicx}
\usepackage{cite}
\usepackage{bm}
\usepackage{comment}
\usepackage{amssymb}
\usepackage{blindtext}
\usepackage{paralist}
\usepackage{enumitem}
\usepackage{xurl}
\usepackage{orcidlink}

\usepackage{titlesec}

\begin{document}

\title{Breaking the HBM Bit Cost Barrier: Domain-Specific ECC for AI Inference Infrastructure}




\author{Rui Xie\orcidlink{0000-0003-3177-5071}, Asad Ul Haq\orcidlink{0009-0003-7975-0102}, Yunhua Fang\orcidlink{0009-0009-4718-8825}, Linsen Ma\orcidlink{0009-0000-8535-7911}, Sanchari Sen\orcidlink{0000-0003-0080-2882}, Swagath Venkataramani\orcidlink{0000-0002-0470-6364}, Liu Liu\orcidlink{0000-0003-0792-8146}, Tong Zhang\orcidlink{0009-0009-8005-0043}

\thanks{Rui Xie, Asad Ul Haq, Yunhua Fang, Linsen Ma, Liu Liu and Tong Zhang are with Rensselaer Polytechnic Institute, Troy, NY 12180 USA.}
\thanks{Sanchari Sen and Swagath Venkataramani are with IBM T.J. Watson Research Center, Yorktown Heights, NY 10598 USA.} 
}



\maketitle
\begin{abstract}
High-Bandwidth Memory (HBM) delivers exceptional bandwidth and energy efficiency for AI workloads, but its high cost per bit, driven in part by stringent on-die reliability requirements, poses a growing barrier to scalable deployment. This work explores a system-level approach to cost reduction by eliminating on-die ECC and shifting all fault management to the memory controller. We introduce a domain-specific ECC framework combining large-codeword Reed--Solomon~(RS) correction with lightweight fine-grained CRC detection, differential parity updates to mitigate write amplification, and tunable protection based on data importance. 
Our evaluation using LLM inference workloads shows that, even under raw HBM bit error rates up to $10^{-3}$, the system retains 78\% of throughput while maintaining at least 97\% PIQA accuracy and 94\% MMLU accuracy relative to error-free HBM.
By treating reliability as a tunable system parameter rather than a fixed hardware constraint, our design opens a new path toward low-cost, high-performance HBM deployment in AI infrastructure.

\end{abstract}


\section{Introduction}
\IEEEPARstart{H}igh-Bandwidth Memory (HBM) has become a foundational element of modern AI computing systems, delivering the high memory bandwidth required to sustain the throughput of large-scale inference and training workloads. However, HBM remains significantly more expensive than conventional DRAM in terms of \$/GB, at least by a factor of 5 to 10~\cite{Koch2024TheMW}, which increasingly threatens the scalability and economic viability of future AI infrastructure. Fundamentally, there are only two viable paths to address this cost challenge: (i) reducing the volume of data stored in HBM through compression, quantization, or algorithmic sparsification~\cite{wan2023efficient}; or (ii) reducing the manufacturing cost of HBM itself. While the first path has attracted extensive research, this work explores the second, approaching it from a system architecture perspective.

The central idea of this work is to reduce HBM cost by {\it substantially} relaxing raw reliability requirements, shifting fault tolerance from HBM stacks to the system level. This principle is well established in storage systems: hard disk drives~(HDDs) and solid-state drives~(SSDs) routinely tolerate elevated raw bit error rate~(BER) by strengthening system-level fault tolerance mechanisms such as error-correcting codes~(ECC)~\cite{zhao2013ldpc}. Unlike these general-purpose media, our focus is domain-specific: HBM deployed for AI inference. Two factors motivate this direction. First, AI inference is expected to dominate both compute and memory usage in future AI infrastructure, making it economically viable to pursue hardware and system-level customization. Second, inference workloads exhibit structural properties, such as sequential access patterns and unequal bit criticality, that expose a large, underexplored design space for HBM ECC solutions. 

Effectively preserving data integrity under relaxed HBM raw reliability requires significantly stronger error correction, and the only viable path is to increase ECC codeword length, from today’s typical 16B or 32B~~\cite{gurumurthi2021hbm3} to much larger blocks such as 512B or 2KB. This is because ECC strength improves exponentially with codeword length. However, deploying large-codeword ECC introduces two critical challenges. First, small random memory accesses, which may still occur despite the predominantly sequential nature of AI inference, can suffer from severe read and write amplification. Second, large-size ECC requires high-complexity decoding hardware, significantly increasing the silicon area and power consumption of the memory controller. To address these limitations, we introduce two complementary techniques: (i) We propose a hybrid ECC architecture inspired by concatenated codes~\cite{forney1965concatenated} that pairs large-size Reed–Solomon (RS) codes with fine-grained Cyclic Redundancy Check (CRC), allowing the system to opportunistically bypass full RS decoding and thereby mitigate read/write amplification. (ii) We can selectively omit ECC protection for low-sensitivity data (e.g., mantissa fields) motivated by unequal error protection~\cite{massey1972variable} to reduce decoding complexity and energy without compromising inference quality. Together, these techniques enable practical adoption of large-codeword ECC in AI inference memory systems, unlocking cost reductions through reliability relaxation. Using large language model (LLM) inference as a case study, our evaluation shows that, even under very high raw HBM bit error rates (e.g., up to $10^{-3}$), the proposed approach can sustain throughput and inference accuracy comparable to systems equipped with ideal, error-free HBM. It is our hope that this work will inspire further research on system- and workload-aware reliability co-design, enabling scalable, cost-effective memory solutions for future AI infrastructure.

\vspace{-7px}
\section{Background and Motivation}
\label{sec:background}
HBM achieves exceptional bandwidth through two key architectural features: a 3D stack of DRAM dies interconnected via dense through-silicon vias~(TSVs), and a wide, high-speed interface that connects directly to the processor die within the same package. This tight integration enables thousands of parallel I/O wires, resulting in aggregate bandwidths of multiple terabytes per second. However, this same integration imposes significant manufacturing complexity and yield sensitivity, driving the cost per bit to at least 5$\sim$10$\times$ higher than conventional DRAM. As AI workloads become increasingly memory-bound, this cost disparity emerges as a critical obstacle to building scalable, cost-efficient AI infrastructure.


A major contributor to HBM cost is the stringent reliability requirement imposed during die testing and binning. In HBM3, each DRAM die integrates short-length on-die ECC at 16B/32B granularity, and the interface permits the host to attach a 2B CRC to each 32B word~\cite{gurumurthi2021hbm3}. The CRC is generated and verified by the host-side controller and transmitted with the data for end-to-end checking. While this combination improves field reliability, it has inherent limits: short codewords cap correction strength and scalability, and both the ECC logic and redundancy are fixed in the DRAM silicon, preventing workload-aware tuning as process scaling increases raw error variability. 

\begin{figure}[htbp]
    \centering
    \includegraphics[width=\linewidth]{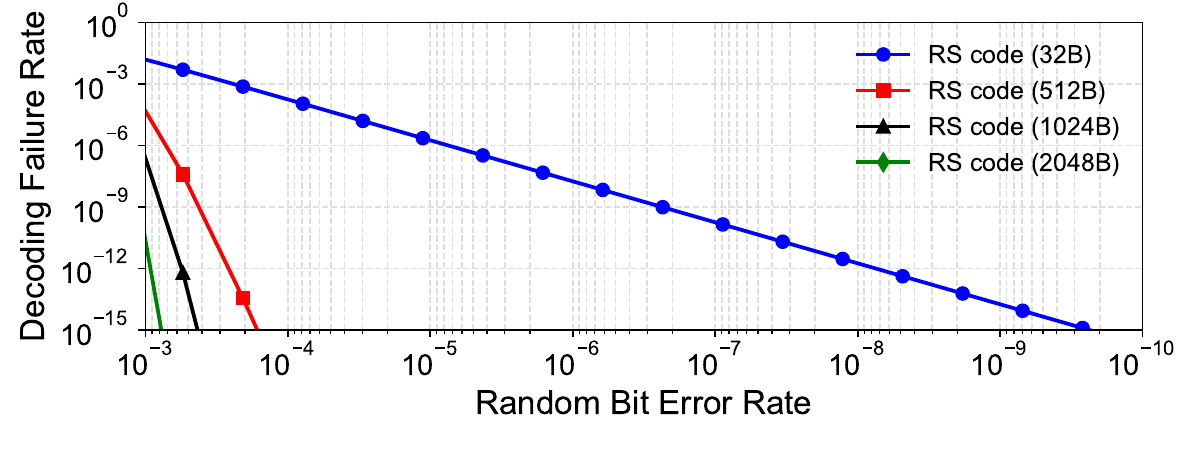}
    \caption{Comparison of decoding failure rate under different codeword sizes with the same code rate of 16/17.}
    \label{fig:decode failure rate}
\end{figure}

To increase yield and reduce manufacturing cost, a more flexible, system-driven reliability model is needed. The core lever is to use \emph{larger ECC codewords} at the same redundancy, which sharply improves correction strength. As a motivating example, Fig.\ref{fig:decode failure rate} models RS codes at a fixed code rate (16/17): starting from 32B (on-die ECC granularity in current HBM~\cite{gurumurthi2021hbm3}), increasing the codeword size to 2KB raises the tolerable raw bit error rate (BER) by more than five orders of magnitude. This illustrates the headroom available from codeword scaling alone, independent of where ECC is implemented.

Implementing such large-codeword ECC directly on DRAM dies is impractical due to limited high-performance logic resources and tight area/power budgets. A practical alternative is to move fault management to the host-side memory controller, which can host high-throughput RS encoders/decoders and support flexible ECC configurations. Critically, the largely sequential access patterns of LLM inference amortize large-codeword overheads in steady state, making controller-based ECC an effective path to relax raw HBM reliability while preserving end-to-end integrity.

However, large-codeword ECC comes with two critical challenges. First, it introduces significant read and write amplification for small or unaligned memory accesses, a scenario still present in AI inference computing devices. Accessing a single corrupted chunk may require fetching or rewriting an entire large-size codeword, inflating latency and bandwidth usage. Second, decoding large ECC blocks requires complex logic, which increases the silicon area and power consumption, especially under multi-TB/s throughput constraint. These challenges define the central question of this work: {\bf Can we unlock the cost-saving potential of large-codeword ECC in HBM-based AI systems while maintaining access efficiency and minimizing controller silicon overhead?} The rest of this paper addresses this question by proposing a domain-specific, controller-centric HBM fault tolerance framework designed to reconcile strong error protection with the performance and efficiency demands of AI inference.

\vspace{-10px}
\section{Proposed Solutions}
\label{sec:method}
\begin{figure}[htbp]
    \centering
    \includegraphics[width=\linewidth]{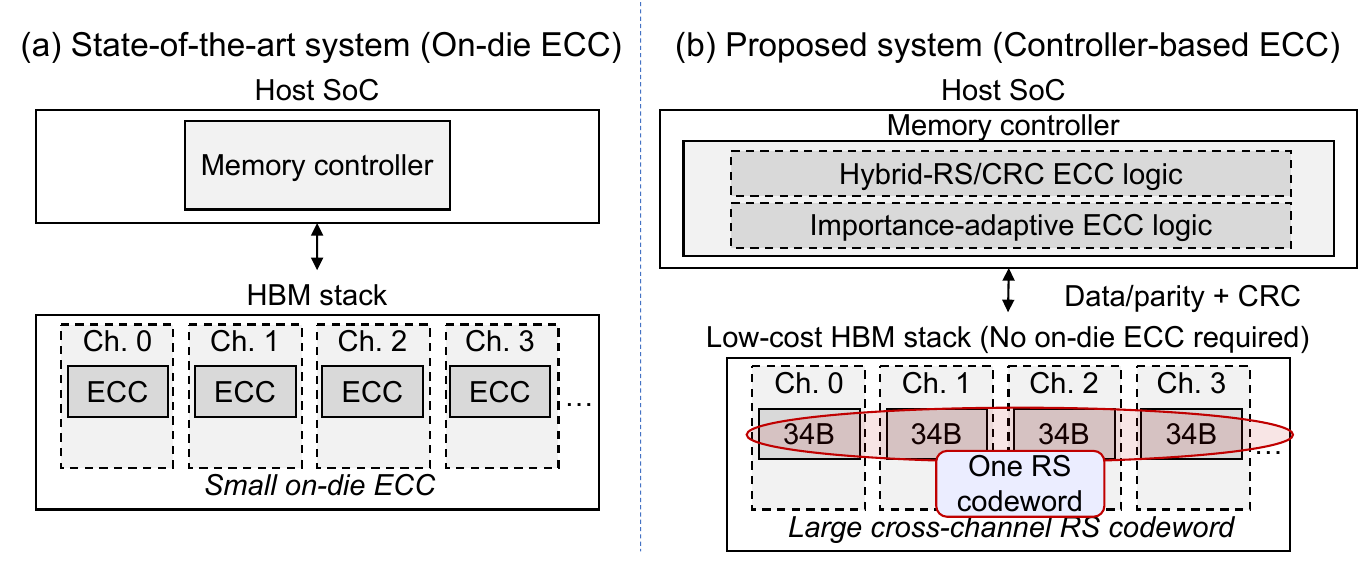}
    \caption{Comparison of HBM ECC Architectures. (a) The state-of-the-art system uses small, independent on-die ECC codewords within each HBM channel. (b) Our proposed system shifts all fault management to the host controller's ECC logic, and use large, cross-channel RS codewords on low-cost HBM stacks.}
    \label{fig:overview}
\end{figure}

This work centers around two simple ideas: (1) applying large-size ECC to substantially relax HBM raw reliability constraints, thereby enabling lower-cost DRAM manufacturing; and (2) leveraging the inherent error resilience of AI inference workloads to reduce the implementation overhead of such ECC. Achieving this requires shifting ECC responsibilities from HBM stacks to host-side memory controller, as illustrated in Fig.~\ref{fig:overview}. 

\subsection{Hybrid ECC Design}
In current HBM products, on-die ECC is implemented using short RS codes, chosen for their robustness to both random and burst errors, and applied to 16B or 32B words. This work preserves the use of RS codes but extends them to operate over much larger codeword sizes~(e.g., 512B or 2KB) under host control. Meanwhile, we also reuse another key feature of the existing HBM interface: the ability for host processors to append a 2B CRC to each 32B data chunk. Instead of relying on CRC solely for basic integrity checking, we integrate it into a hybrid RS/CRC scheme, where per-chunk CRCs serve as low-latency filters that detect errors and avoid triggering full RS decoding. This opportunistic use of CRC dramatically reduces read and write amplification, without modifying the HBM physical interface or access granularity.

To organize large-size RS codeword efficiently, we exploit the inherent parallelism of HBM stacks (e.g., each HBM3E stack includes 16 independent channels, each 128 bits wide). We stripe each RS codeword across $s$ such channels to maximize bandwidth and parallel access, as illustrated in Fig.~\ref{fig:overview}. Specifically, each RS codeword contains a number of 32B user data chunks and one or multiple 32B parity chunks. Each 32B chunk, whether data or parity, is appended with its own 2B CRC, forming a 34B unit. These  CRC-augmented chunks are striped sequentially across the 
$s$ channels. Because each chunk is self-contained with its own CRC, this structure supports fine-grained error detection with minimal latency and bandwidth overhead. 
This hybrid scheme requires no change to the HBM physical interface or access granularity.

\begin{figure}[htbp]
    \centering
    \includegraphics[width=\linewidth]{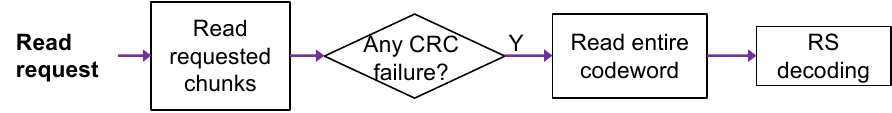}
    \caption{Operational flow of serving a random read request.}
    \label{fig:read_flow}
\end{figure}

\noindent\textbf{Random Read}: Fig.~\ref{fig:read_flow} illustrates the procedure for handling random read requests. To serve a random read that spans $k$ 32B data chunks within a single RS codeword, the memory controller retrieves all $k$ chunks and their associated 2B CRCs from HBM. It then verifies all CRCs locally. If \emph{all} CRCs pass, the corresponding data chunks are immediately returned to the host. However, if \emph{any} CRC check fails, the controller triggers an \emph{escalation} procedure: It fetches the remaining chunks in the RS codeword from HBM and performs RS decoding to recover all the $k$ data chunks. Let $p$ denote the HBM raw BER, we can calculate the probability of triggering full RS codeword fetching/decoding as $P_{dec}=1-(1-p)^{272\cdot k}$. For example, under the raw BER of as high as $1\times10^{-4}$, $P_{dec}$ is 2.7\% and 10.3\% when $k$ is 1 and 4, respectively. 

\begin{figure}[htbp]
    \centering
    \includegraphics[width=\linewidth]{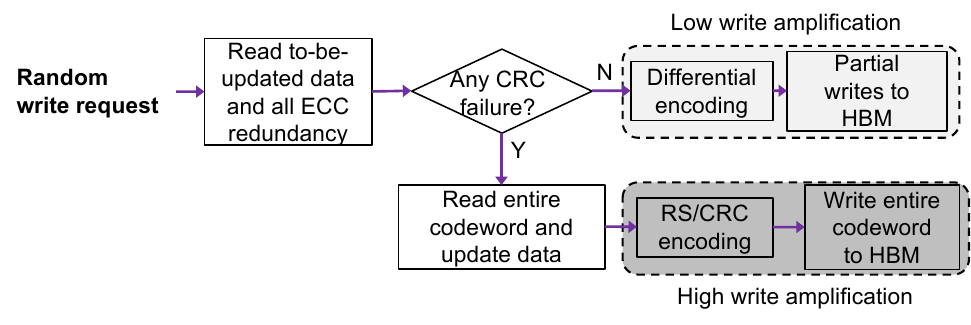}
    \caption{Operational flow of serving a random write request.}
    \label{fig:write_flow}
\end{figure}

\noindent\textbf{Random Write}: Fig.~\ref{fig:write_flow} illustrates the procedure for handling random write requests. For random writes that update $k$ 32B chunks within a single RS codeword, the memory controller performs a CRC-guided opportunistic partial update to avoid full read-modify-write operations. First, it fetches the current version of the $k$ target chunks along with all the $r$ RS parity chunks. If \emph{all} CRCs on the fetched chunks pass, the controller proceeds with a partial update as follows. Assume each RS codeword protects a total of $m$ 32B chunks. We construct two $m \times$32B vectors ${\bf D}_{old}$ and ${\bf D}_{new}$: one containing the old values of the $k$ chunks in their original positions and zeros elsewhere, and another containing the new values of those same $k$ chunks (with zeros in the remaining positions). These two sparse vectors are independently RS-encoded to generate two sets of $r$ parity chunks $\text{RS}({\bf D}_{old})$ and $\text{RS}({\bf D}_{new})$. Because RS code is a linear block code, the difference between them captures the net parity update due to the modified data chunks. Accordingly, the new RS parity ${\bf D}_{new}$ is computed by XORing the original parity ${\bf P}_{old}$ with the difference of the two RS-encoded results:
$\mathbf{P}_{\text{new}} = \mathbf{P}_{\text{old}} \oplus \text{RS}(\mathbf{D}_{\text{new}}) \oplus \text{RS}(\mathbf{D}_{\text{old}})$.

The controller then writes the updated data chunks and the updated parity back to HBM. However, if any CRC check fails during the initial read, the controller escalates to a full codeword fetch and re-encodes the entire word, falling back to a conventional read-modify-write path as illustrated in Fig.~\ref{fig:write_flow}.

\noindent\textbf{Sequential Read/Write}: Sequential read and write requests are handled in a much simpler manner. For sequential reads, we directly fetch the entire RS codeword, including data and parity, and perform RS decoding without first checking CRCs. At high BER, errors are likely, and skipping CRC avoids redundant checks. The RS decoder can naturally early-terminate if no errors are present, minimizing latency and energy. For sequential writes, the controller performs RS encoding and per-chunk CRC encoding in a single pass, then writes the full hybrid RS/CRC codeword directly to HBM. 
 
\vspace{-10px}
\subsection{Importance‑Adaptive ECC Protection}\label{sec:adaptive}

As discussed above, implementing large-codeword RS decoders incurs substantial silicon cost, particularly under multi-TB/s HBM throughput constraints. To mitigate this, we use \emph{importance-adaptive ECC} that selectively protects only the data bits that matter most for inference quality. Many AI inference workloads, especially with BF16 or FP8, are numerically resilient: mantissa flips often have negligible effect compared to exponent flips. By applying ECC only to the sensitive portions, we cut RS decoder load, ECC traffic, and energy. This mechanism is \textit{orthogonal} to numeric format: it operates on bit-planes and only requires a per-format criticality map. We report BF16 in experiments for space and prevalence.

To support this functionality efficiently, we adopt a bit-plane–oriented in-memory data placement scheme. Consider a block of $m$ numerical values $\{x_1, \dots, x_m\}$, each $n$ bits wide. Each value $x_j$ is represented as a bit vector $[b_{j,n-1}, b_{j,n-2}, \dots, b_{j,0}]$, where $b_{j,i} \in \{0,1\}$ is the $i$-th bit of $x_j$. The $i$-th \emph{bit-plane} is defined as $P_i = \{b_{1,i}, b_{2,i}, \dots, b_{m,i}\}$, i.e., all bits at position $i$ across the block. Let $\mathcal{S} \subseteq \{0, \dots, n-1\}$ denote the set of \emph{critical} planes that require protection, and let $\gamma = \frac{|\mathcal{S}|}{n}$ denote the \emph{protected-plane ratio}, with $0 \le \gamma \le 1$. Only bits in planes $\mathcal{S}$ are processed through CRC and RS encoding; all other bits bypass the ECC logic entirely. Accordingly, we can reduce the RS decoder silicon cost by approximately $1-\gamma$.

By integrating importance-adaptive ECC with chunk-level CRC filtering and incremental parity updates, we construct a tunable fault-tolerance framework. By adjusting $\gamma$ to match workload-specific error sensitivity, system designers can trade off memory controller complexity and energy consumption against yield improvements in HBM manufacturing, thereby reducing the system cost without compromising AI inference accuracy.

\vspace{-10px}
\section{Evaluation}
We use Intel Pin~\cite{intel_pin_tool} to collect fine-grained LLM inference traces and drive a hybrid evaluation composed of (i) an analytic ECC model (CRC pass probability, RS escalations/read-modify-write amplification, effective bandwidth) and (ii) a modified DRAMSim3~\cite{li2020dramsim3} configured as an HBM3-class system with \(16\times 128\)-bit channels and \(\approx\)1~TB/s, augmented with hooks for ECC-induced traffic and parameterized encoder/decoder service times. 
\subsection{Impact of Codeword Size on Inference Throughput}

\begin{figure}[htbp]
  \centering
  \includegraphics[width=.95\linewidth]{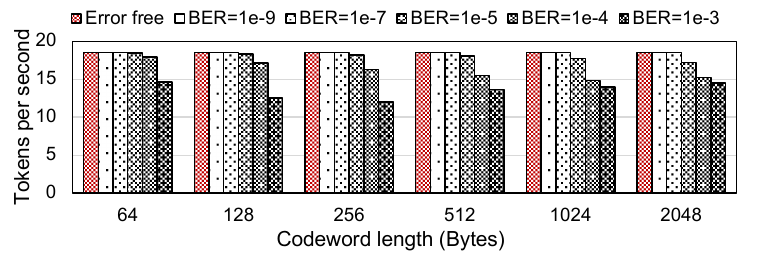}
  \caption{Inference throughput vs. RS codeword length under various BERs.}
  \label{fig:throughput}
\end{figure}

Fig.~\ref{fig:throughput} shows LLM inference throughput (tokens/s) as a function of RS codeword length, ranging from 64B to 2048B, under six different raw HBM BERs. Experiments were conducted on the DeepSeek R1 670B model (with 10\% active weights), using traces sampled by the Pin tool~\cite{intel_pin_tool}, assuming a 1~TB/s HBM under 99\% sequential and 1\% random accesses. Under error-free scenario and at BER of $10^{-9}$, throughput remains constant at 18.51 tokens/s for all codeword sizes, since CRC (almost) never fails and each 32B user chunk maps directly to a 34B transfer unit, making codeword length irrelevant to the performance. As BER increases to $10^{-7}$, CRC begins to occasionally fail, with a per-chunk failure probability of approximately $2.7 \times 10^{-5}$. However, the impact on throughput is negligible, decreasing only slightly from 18.51 tokens/s (64B) to 18.49 tokens/s (2048B). At BER of $10^{-5}$, the higher failure rate leads to more frequent RS decoding, and throughput declines steadily from 18.44 to 17.20 tokens/s as codeword length increases. At BER of $10^{-4}$, the trend becomes non-monotonic: throughput first drops to 14.85 tokens/s at 1024B due to random access amplification, but then recovers slightly at 2048B (15.21 tokens/s), where sequential access benefits begin to dominate. At BER of $10^{-3}$, the initial decline is sharper, dropping to 12.05 tokens/s at 256B, followed by recovery at longer codewords (e.g., 14.51 tokens/s at 2048B). These results demonstrate that even under severe error conditions, the proposed ECC design sustains a substantial fraction of baseline throughput, reaching over 78\% of the ideal performance at the worst tested BER.


\vspace{-10px}

\vspace{-15px}
\subsection{Impact of Random Access Intensity}
\begin{figure}[htbp]
  \centering
  \includegraphics[width=.95\linewidth]{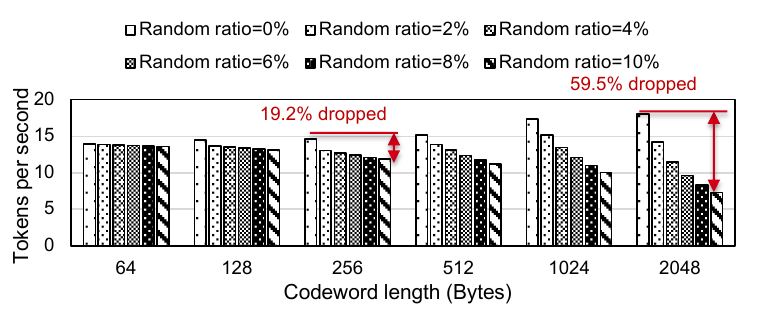}
  \caption{Inference throughput vs.\ random access ratio at BER of \(10^{-3}\) under codeword length from 64B to 2048B.}
  \label{fig:rand_intensity}
\end{figure}

We sweep the random‐access ratio from 0\% to 10\% at BER $10^{-3}$ (Fig.~\ref{fig:rand_intensity}). With purely sequential traffic, throughput rises with codeword size from 13.90 tokens/s (64B) to 18.05 tokens/s (2048B) due to parity amortization. Even small randomness erodes this benefit: at 2\% random, 2048B drops to 14.26 tokens/s (-21.0\%) while 64B is 13.85 tokens/s (-0.4\%). At 10\% random, 64B holds 13.64 tokens/s (-1.9\%), 256B falls to 11.87 tokens/s (-19.2\%), and 2048B collapses to 7.31 tokens/s (-59.5\%). The cause is CRC failures on 34B chunks that escalate into full-codeword fetches or read-modify-writes, whose penalty grows with codeword size. For workloads with even modest randomness, moderate codeword sizes (256B–512B) strike the best balance between parity amortization and escalation overhead.

\vspace{-12px}

\subsection{Importance-Adaptive Protection}

\begin{figure}[htbp]
  \centering
  \includegraphics[width=.95\linewidth]{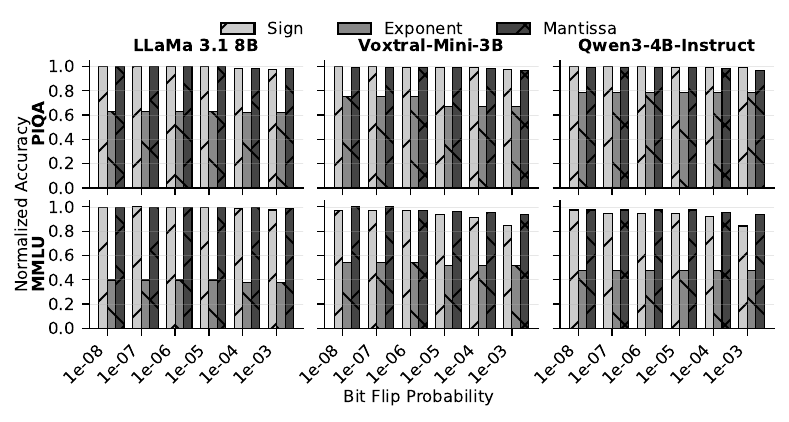}
\caption{Motivational study demonstrating the critical vulnerability of exponent bits in unprotected BF16 models. Accuracy is evaluated on PIQA (top) and MMLU (bottom) after injecting bit-flips into the sign, exponent, and mantissa.}
  \label{fig:bitfield_sensitivity}
\end{figure}

Our importance-adaptive protection scheme is inspired by the \emph{motivational} study in Fig.~\ref{fig:bitfield_sensitivity}, which reports normalized accuracy under targeted BF16 bit flips for LLaMA-3.1-8B~\cite{llama-3-1-8b-instruct}, Voxtral-Mini-3B~\cite{Voxtral-Mini-3B-2507}, and Qwen3-4B~\cite{qwen3-4b-instruct-2507} on PIQA~\cite{bisk2020piqa} and MMLU~\cite{hendrycks2020measuring} as we sweep flip rates from $10^{-8}$ to $10^{-3}$. This is a stress test on \emph{unprotected} bit-planes to rank vulnerability.
The trend is consistent: exponent flips are the dominant failure mode, while sign and mantissa flips have much smaller impact. 
At $10^{-3}$ with exponent-only corruption, PIQA drops to a worst case of 61.7\% and MMLU to 38.1\%. In contrast, if we protect the exponent and allow faults only in sign or mantissa, accuracy remains high even at $10^{-3}$ in the worst case: on PIQA it is at least 97.3\% (sign) and 96.8\% (mantissa). On MMLU it is at least 84.2\% (sign) and 94.0\% (mantissa).
These results motivate an importance-adaptive policy that always protects exponent bit-planes and uses reduced or no protection for sign/mantissa, cutting ECC load with minimal accuracy loss. We show BF16 here for space, the mechanism is format-agnostic.

\vspace{-10px}

\begin{figure}[htbp]
  \centering
  \includegraphics[width=.93\linewidth]{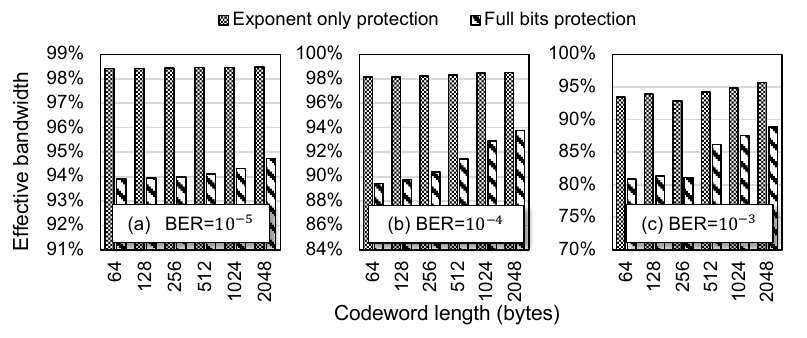}
  \caption{Effective HBM bandwidth utilization on LLaMA-3.1-8B (MMLU) for exponent-only vs. full-bit ECC under BERs of $10^{-5}$, $10^{-4}$, and $10^{-3}$ across codeword lengths.}
  \label{fig:eff_bw_exp}
\end{figure}

Fig.~\ref{fig:eff_bw_exp} explores the efficiency gains of concentrating ECC protection solely on the exponent field (on LLaMA-3.1-8B and MMLU). We compare effective HBM bandwidth utilization between exponent-only and full-bit ECC strategies across RS codeword lengths ranging from 64B to 2048B, under three increasingly aggressive BERs ($10^{-5}$, $10^{-4}$, and $10^{-3}$). At $10^{-5}$, exponent-only ECC achieves 98.4\% utilization at 64B and 98.5\% at 2048B, while full-bit ECC reaches only 93.9\% and 94.7\%, respectively. As the BER rises to $10^{-4}$, the efficiency gap widens further. For instance, with 256B codewords, exponent-only ECC maintains 98.23\% bandwidth, compared to 90.4\% for full-bit ECC. Even under extreme noise at $10^{-3}$, exponent-only protection holds up: 93.5\% at 64B and 95.7\% at 2048B, whereas full-bit ECC drops to 80.9\% and 89.0\%, respectively. Across all settings, adaptive protection consistently reduces ECC-related traffic overhead, with bandwidth improvements reaching up to 12.6\%. 

Therefore, in practice, we always protect the exponent planes. Together these results demonstrate the effectiveness of importance-aware ECC in balancing robustness and efficiency. By aligning protection strength with data sensitivity, focusing on the exponent field, this approach reduces bandwidth consumption and decoder workload, while preserving accuracy under high error rates. Importance-adaptive ECC thus offers a compelling direction for lowering HBM cost and controller complexity, opening the door to more scalable and resilient AI memory systems.

\vspace{-10px}

\vspace{-20px}
\section{Conclusion}
\label{sec:conclusion}
This work proposes a domain-specific ECC framework that enables a system-level approach to reducing HBM cost for AI inference. By eliminating on-die ECC and shifting error management to the memory controller, we demonstrate that strong fault tolerance can be maintained even under elevated error rates. Our design combines large-codeword RS correction with lightweight CRC filtering, differential parity updates, and importance-adaptive protection to balance reliability and efficiency. Evaluation on LLM workloads shows that this approach sustains performance and accuracy while significantly relaxing raw HBM reliability. By treating reliability as a tunable, workload-aware parameter, our work opens new directions for scalable, cost-effective memory system design in AI infrastructure.

\bibliographystyle{IEEEtran}
\bibliography{ref}

\end{document}